\documentclass[sigconf]{acmart}

\usepackage{booktabs} % For formal tables

\usepackage{balance}
\usepackage[ruled]{algorithm2e} % For algorithms
\usepackage{amssymb}
\usepackage{amsmath}

\SetAlFnt{\small}
\SetAlCapFnt{\small}
\SetAlCapNameFnt{\small}
\SetAlCapHSkip{0pt}
\IncMargin{-\parindent}

\setcopyright{none}
\renewcommand\footnotetextcopyrightpermission[1]{} 
\copyrightyear{2018} 
\acmYear{2018} 
\acmConference[MM '18]{2018 ACM Multimedia Conference}{October 22--26, 2018}{Seoul, Republic of Korea}
\acmBooktitle{2018 ACM Multimedia Conference (MM '18), October 22--26, 2018, Seoul, Republic of Korea}
\acmPrice{15.00}
\acmDOI{10.1145/3240508.3240712}
\acmISBN{978-1-4503-5665-7/18/10}

\begin{document}

\fancyhead{}
% Title portion
\title{ Webly Supervised Joint Embedding for Cross-Modal \\ Image-Text Retrieval}
%\titlenote{We can add a note to the title}

\author{Niluthpol Chowdhury Mithun}
\affiliation{%
  \institution{University of California, Riverside, CA}
}
\email{nmith001@ucr.edu}

\author{Rameswar Panda}
\affiliation{%
  \institution{University of California, Riverside, CA}
}
\email{rpand002@ucr.edu}

\author{Evangelos E. Papalexakis}
\affiliation{%
  \institution{University of California, Riverside, CA}
}
\email{epapalex@cs.ucr.edu}

\author{Amit K. Roy-Chowdhury}
\affiliation{%
  \institution{University of California, Riverside, CA}
}
\email{amitrc@ece.ucr.edu}

\begin{abstract}
Cross-modal retrieval between visual data and natural language description remains a long-standing challenge in multimedia. While recent image-text retrieval methods offer great promise by learning deep representations aligned across modalities, most of these methods are plagued by the issue of training with small-scale datasets covering a limited number of images with ground-truth sentences. Moreover, it is extremely expensive to create a larger dataset by annotating millions of images with sentences and may lead to a biased model. Inspired by the recent success of webly supervised learning in deep neural networks, we capitalize on readily-available web images with noisy annotations to learn robust image-text joint representation. Specifically, our main idea is to leverage web images and corresponding tags, along with fully annotated datasets, in training for learning the visual-semantic joint embedding. We propose a two-stage approach for the task that can augment a typical supervised pair-wise ranking loss based formulation with weakly-annotated web images to learn a more robust visual-semantic embedding. Experiments on two standard benchmark datasets demonstrate that our method achieves a significant performance gain in image-text retrieval compared to state-of-the-art approaches.

\end{abstract}



\maketitle

\renewcommand{\shortauthors}{N. Mithun et al.}

\section{Introduction}

\begin{figure}[t]
\vspace{0.3cm}
\centering
	\includegraphics[width=0.48\textwidth]{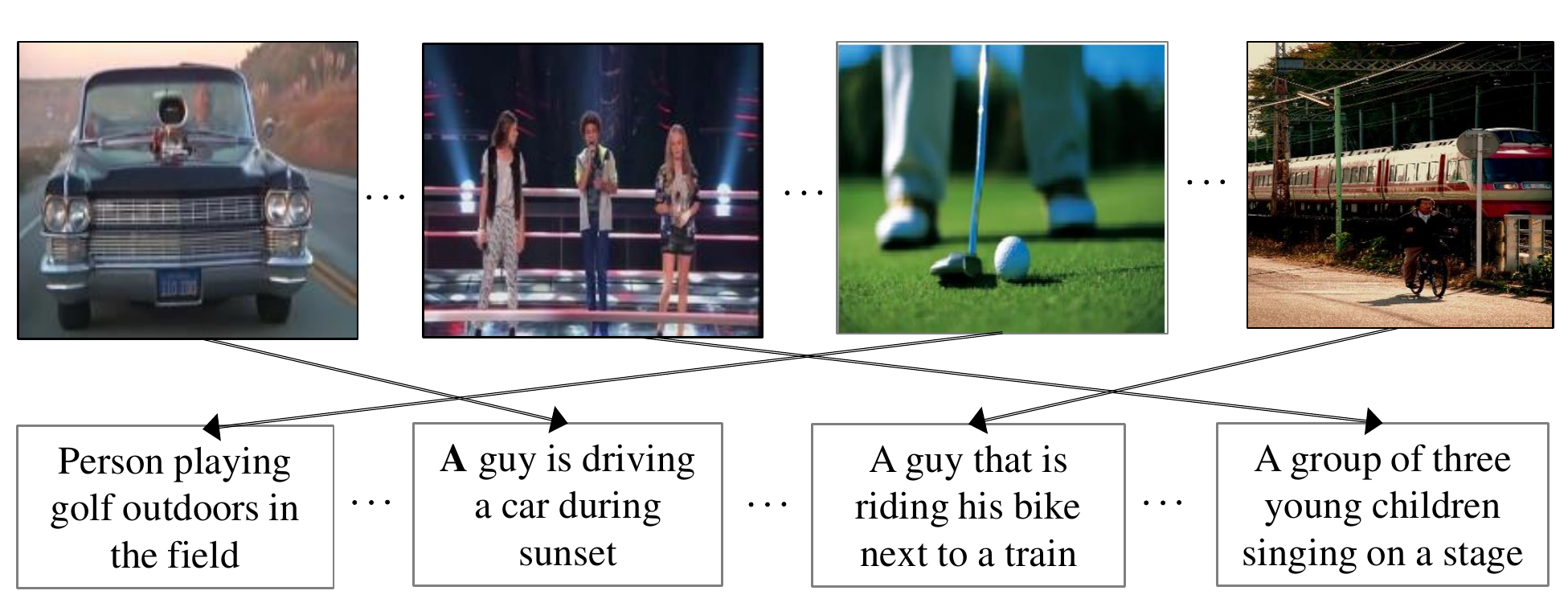}
	\caption{Illustration of image-text retrieval task: Given a text query, retrieve and rank images from the database based on how well they depict the text or vice versa.}
	\vspace{-0.4cm}
	\label{fig:problem}
\end{figure}

Joint embeddings have been widely used in multimedia data mining as they enable us to integrate the understanding of different modalities together. These embeddings are usually learned by mapping inputs from two or more distinct domains (e.g., images and text) into a common latent space, where the transformed vectors of semantically associated inputs should be close. Learning an appropriate embedding is crucial for achieving high-performance in many multimedia applications involving multiple modalities.
%a cross-modal retrieval task. 
In this work, we focus on the task of cross-modal retrieval between images and language (See Fig.~\ref{fig:problem}), i.e., the retrieval of images given sentence query, and retrieval of text from a query image.  

The majority of the success in image-text retrieval task has been achieved by the joint embedding models trained in a supervised way using image-text pairs from hand-labeled image datasets (e.g., MSCOCO \cite{chen2015microsoft}, Flickr30k\cite{plummer2015flickr30k}). Although, these datasets cover a significant number of images (e.g., about 80k in MSCOCO and 30K in Flickr30K), creating a larger dataset with image-sentence pairs is extremely difficult and labor-intensive \cite{krause2016unreasonable}. Moreover, it is generally feasible to have only a limited number of users to annotate training images, which may lead to a biased model \cite{van2016stereotyping, hu2018exploring, zhao2017men}. Hence, while these datasets provide a convenient modeling assumption, they are very restrictive considering the enormous amount of rich descriptions that a human can compose \cite{karpathy2015deep}. Accordingly, although trained models show good performance on benchmark datasets for image-text retrieval task, applying such models in the open-world setting is unlikely to show satisfactory cross-dataset generalization (training on a dataset, testing on a different dataset) performance.

\begin{figure*}[t]
%\vspace{0.1cm}
\centering
	\includegraphics[width=1\textwidth]{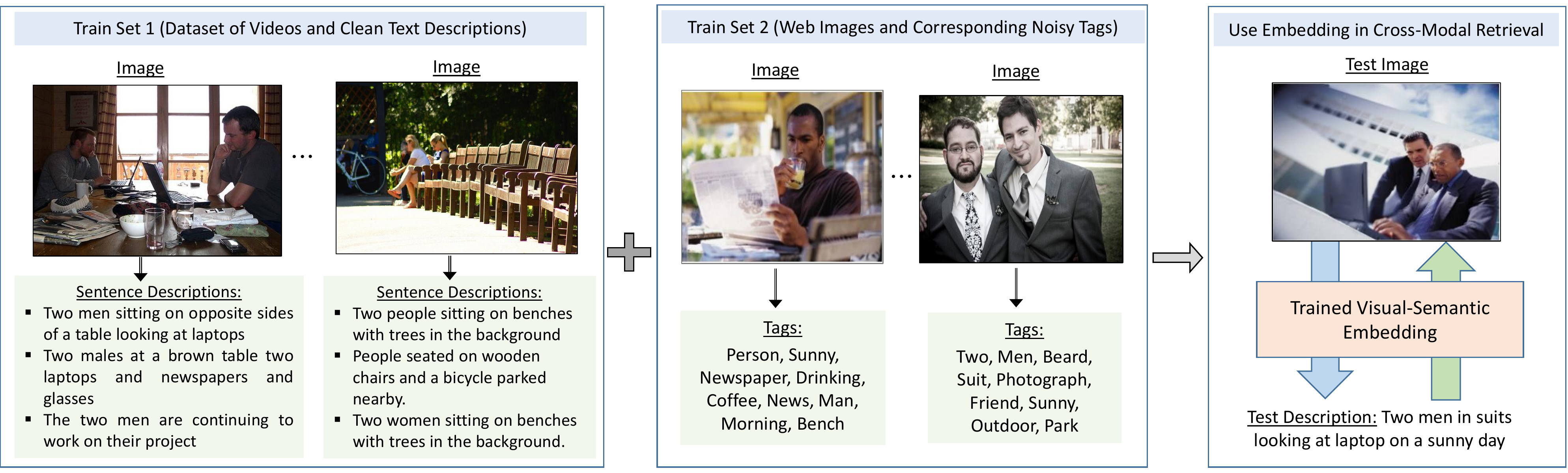}
	\caption{The problem setting of our paper. Our goal is to utilize web images associated with noisy tags to learn a robust visual-semantic embedding from a dataset of clean images with ground truth sentences. We test the learned latent space by projecting images and text descriptions from the test set in the embedding and perform cross-modal retrieval.}
	\label{fig:overview}
	\vspace{-0.1cm}
\end{figure*}

On the other hand, streams of images with noisy tags are readily available in datasets, such as Flickr-1M \cite{huiskes2008mir}, as well as in nearly infinite numbers on the web. Developing a practical system for image-text retrieval considering a large number of web images is more likely to be robust. However, inefficient utilization of weakly-annotated images may increase ambiguity and degrade performance. Motivated by this observation, we pose an important question in this paper: \textit{Can a large number of web images with noisy annotations be leveraged upon with a fully annotated dataset of images with textual descriptions to learn better joint embeddings?} Fig.~\ref{fig:overview} shows an illustration of this scenario. This is an extremely relevant problem to address due to the difficulty and non-scalability of obtaining a large amount of human-annotated training set of image-text pairs. 

In this work, we study how to judiciously utilize web images to develop a successful image-text retrieval system. We propose a novel framework that can augment any ranking loss based supervised formulation with weakly-supervised web data for learning robust joint embeddings. Our approach consistently outperforms previous approaches significantly in cross-modal image-text retrieval tasks. We believe our efforts will provide insights to the researchers working in this area to focus on the importance of large scale web data for efficiently learning a more comprehensive representation from multimodal data.

\vspace{-0.2cm}
\subsection{Overview of the Proposed Approach } 
In the cross-modal image-text retrieval task, an embedding network is learned to project image features and text features into the same joint space, and then the retrieval is performed by searching the nearest neighbor in the latent space. In this work, we attempt to utilize web images annotated with noisy tags for improving joint embeddings trained using a dataset of images and ground-truth sentence descriptions. However, combining web image-tag pairs with image-text pairs in training the embedding is non-trivial. The greatest obstacle arises from noisy tags and the intrinsic difference between the representation of sentence description and tags. A typical representation of text is similar to, and yet very different from the representation of tags. Sentences are usually represented using RNN-based encoder with word-to-vec (Word2Vec) model, providing sequential input vectors to the encoder. In contrast, tags do not have sequential information and a useful representation of tags can be tf-idf weighted BOW vectors or the average of all Word2Vec vectors corresponding to the tags. 

\begin{figure*}[t]
\centering
	\includegraphics[width=0.98\textwidth]{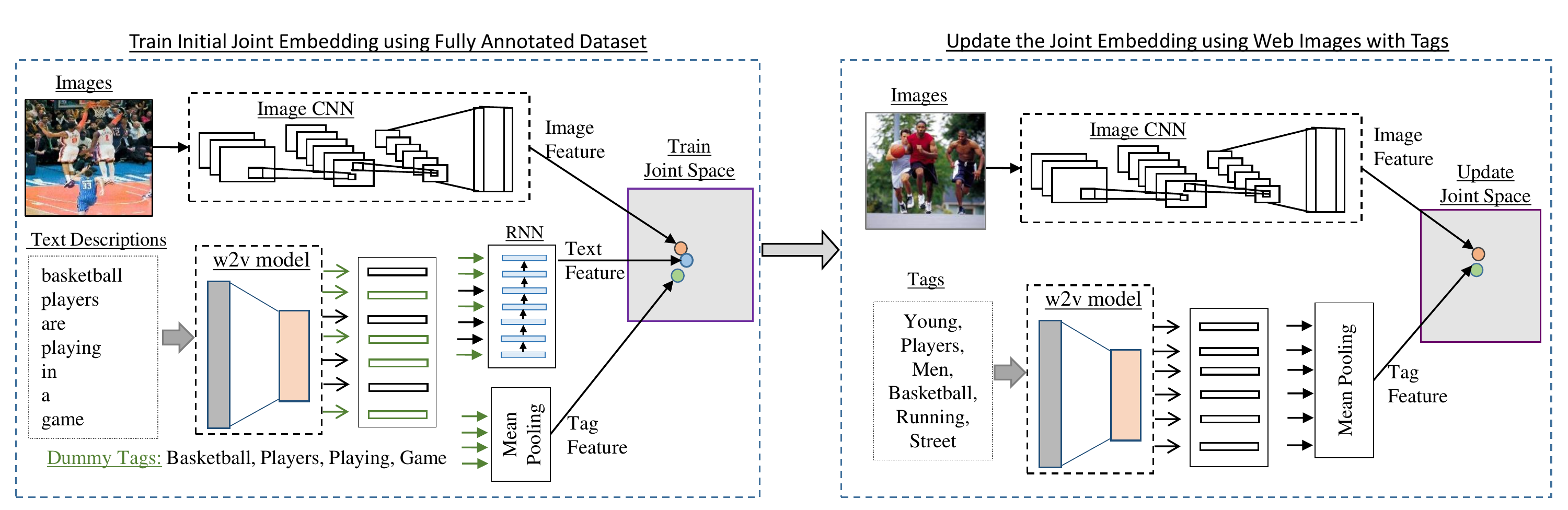}	\caption{A brief illustration of our proposed framework for learning visual-semantic embedding model utilizing image-text pairs from a dataset and image-tag pairs from the web. First, a dataset of images and their sentence descriptions are used to learn an aligned image-text representation. Then, we update the joint representation using web images and corresponding tags. The trained embedding is used in image-text retrieval task. Please see Section \ref{methods} for details.}
	\label{fig:block_diagram}
	\vspace{-0.2cm}
\end{figure*}

To bridge this gap, we propose a two-stage approach that learns the joint image-text representation. Firstly, we use a supervised formulation that leverages the available clean image-text pairs from a dataset to learn an aligned representation that can be shared across three modalities (e.g., image, tag, text). As tags are not available directly in the datasets, we consider nouns and verbs from a sentence as dummy tags (Fig.~\ref{fig:block_diagram}). We leverage ranking loss based formulation with image-text and image-tags pairs to learn a shared representation across modalities. Secondly, we utilize weakly-annotated image-tags pairs from the web (e.g., Flickr) to update the previously learned shared representation, which allows us to transfer knowledge from thousands of freely available weakly annotated images to develop a better cross-modal retrieval system. Our proposed approach is also motivated by learning using privileged information (LUPI) paradigm \cite{vapnik2009new, sharmanska2013learning} and multi-task learning strategies in deep neural networks \cite{ruder2017overview, bingel2017identifying} that share representations between closely related tasks for enhanced learning performance.

\subsection{Contributions} We address a novel and practical problem in this paper---how to exploit large-scale web data for learning an effective multi-modal embedding without requiring a large amount of human-crafted training data. Towards solving this problem, we make the following main contributions.

$\bullet$ 
We propose a webly-supervised approach utilizing web image collection with associated noisy tags, and a clean dataset containing images and ground truth sentence descriptions for learning robust joint representations.

$\bullet$ We develop a novel framework with pair-wise ranking loss for augmenting a typical supervised method with weakly-supervised web data to learn a more robust joint embedding.

$\bullet$ We demonstrate clear performance improvement in image-text retrieval task using proposed web-supervised approach on Flickr30K \cite{plummer2015flickr30k} and MSCOCO datasets \cite{lin2014microsoft}.

%-------------------------------------------------------------------------
\section{Related Work}

\vspace{0.1cm}
\textbf{Visual-Semantic Embedding: } Joint visual-semantic models have shown excellent performance on several multimedia tasks, e.g., cross-modal retrieval \cite{wang2016learning, klein2015associating, huang2017learning, mithun2018learning}, image captioning \cite{mao2014deep,karpathy2015deep}, image classification \cite{hubert2017learning, frome2013devise, gong2014multi} video summarization \cite{choi2017textually,plummer2017enhancing}. Cross-modal retrieval methods require computing semantic similarity between two different modalities, i.e., vision and language. Learning joint visual-semantic representation naturally fits to our task of image-text retrieval since it is possible to directly compare visual data and sentence descriptions in such a joint space~\cite{FaghriFKF17, nam2016dual}. 

%\textcolor{red}{Amit: One sentence on how this work is similar or different.}

\vspace{0.1cm}
\textbf{Image-Text Retrieval: } Recently, there has been significant interest in developing powerful image-text retrieval methods in multimedia, computer vision and machine learning communities \cite{karpathy2014deep, henning2017estimating}.
In \cite{farhadi2010every}, a method for mapping visual and textual data to a common space based on extracting a triplet of object, action, and scene is presented. A number of image-text embedding approaches has been developed based on Canonical Correlation Analysis (CCA) \cite{yan2015deep, socher2010connecting, hodosh2013framing, gong2014multi}. Ranking loss has been used for training the embedding in most recent works relating image and language modality for image-text retrieval \cite{kiros2014unifying, frome2013devise, wang2017learning, FaghriFKF17, nam2016dual}. In \cite{frome2013devise}, words and images are projected to a common space utilizing a ranking loss that applies a penalty when an incorrect label is ranked higher than the correct one. A bi-directional ranking loss based formulation is used to project image features and sentence features to a joint space for cross-modal image-text retrieval in \cite{kiros2014unifying}. 

Several image-text retrieval methods extended this work \cite{kiros2014unifying} with slight modifications in the loss function \cite{FaghriFKF17}, similarity calculation \cite{vendrov2015order, wang2017learning} or input features \cite{nam2016dual}. In~\cite{FaghriFKF17}, the authors modified the ranking loss based on violations incurred by relatively hard negatives and is the current state-of-the art in image-text retrieval task. An embedding network is proposed in \cite{wang2017learning} that uses the bi-directional ranking loss along with neighborhood constraints. Multi-modal attention mechanism is proposed in \cite{nam2016dual} to selectively attend to specific image regions and sentence fragments and calculate similarity. A multi-modal LSTM network is proposed in \cite{huang2017instance} that recurrently select salient pairwise instances from image and text, and aggregate local similarity measurement for image-sentence matching. Our method complements the works that project words and images to a common space utilizing a bi-directional ranking loss. The proposed formulation could be extended and applied to most of these approaches with little modifications.

\vspace{0.1cm}
\textbf{Webly Supervised Learning: } The method of manually annotating images for training does not scale well to the open-world setting as it is impracticable to collect and annotate images for all relevant concepts \cite{li2017learning, mithun2016generating}. Moreover, there exists different types of bias in the existing datasets \cite{van2016stereotyping, torralba2011unbiased, khosla2012undoing}. In order to circumvent these issues, several recent studies focused on using web images and associated metadata as auxiliary source of information to train their models \cite{li2017attention, gong2017multimodal, sun2015temporal}. Although web images are noisy, utilizing such weakly-labeled images has been shown to be very effective in many multimedia tasks \cite{gong2014improving, li2017attention, joulin2016learning}  

Our work is motivated by these works on learning more powerful models by realizing the potential of web data. As the largest MSCOCO dataset for image-sentence retrieval has only 80K training images, we believe it is extremely crucial and practical to complement scarcer clean image-sentence data with web images to improve the generalization ability of image-text embedding models. Most relevant to our work is \cite{gong2014improving}, where authors constructed a dictionary by taking a few thousand most common words and represent text as tf-idf weighted bag of words (BoW) vectors that ignore word order and represents each caption as a vector of word frequencies. Although, such a textual feature representation allows them to utilize the same feature extractor for sentences and set of tags, it fails to consider the inherent sequential nature present in sentences in training image-sentence embedding models.

%\textcolor{red}{Amit: One sentence on how this work is similar or different. Something like we take a similar approach to enrich the joint embedding models. In related work, it is not enough to just cite a list of papers. You should also mention their relevance to your paper. One sentence for each para is enough, but something needs to be said.}
%\reminder{V: I agree, and especially with respect to the Webly Supervised stuff, we may want to clearly outline what is our novelty so that reviewers won't have to think and make wrong assumptions.}

%\textcolor{red}{Amit: Read up to here.}

%-------------------------------------------------------------------------
\section{Approach}  \label{methods}

In this section, we first describe the network structure (Section~\ref{feature}). Then, we revisit the basic framework for learning image text mapping using pair-wise ranking loss (Section~\ref{vse}). Finally, we present our proposed strategy to incorporate the tags in the framework to learn an improved embedding (Section~\ref{trimodal}).

\subsection{Network Structure and Input Feature} \label{feature}

\vspace{0.2cm}
\textbf{Network Structure:} We learn our joint embedding model using a deep neural network framework. As shown in Fig.~\ref{fig:block_diagram}, our model has three different branches for utilizing image, sentence, and tags. Each branch has different expert network for a specific modality followed by two fully connected embedding layers. The idea is that the expert networks will focus on identifying modality-specific features at first and the embedding layers will convert the modality-specific features to modality-robust features. The parameters of these expert networks can be fine-tuned together with training the embedding layers. For simplicity, we keep image encoder (e.g., pre-trained CNN) and tag encoder (e.g., pre-trained Word2Vec model) fixed in this work. The word embedding and the GRU for sentence representation are trained end-to-end.

\vspace{0.1cm}
\textbf{Text Representation:} For encoding sentences, we use Gated Recurrent Units (GRU) \cite{chung2014empirical}, which has been used for representing sentences in many recent works \cite{FaghriFKF17, kiros2014unifying}. We set the dimensionality of the joint embedding space, $D$, to 1024. The dimensionality of the word embeddings that are input to the GRU is 300. 

\vspace{0.1cm}
\textbf{Image Representation:} For encoding image, we adopt a deep CNN model trained on ImageNet dataset as the encoder. Specifically, we experiment with state-of-the-art 152 layer ResNet model \cite{he2016deep} and 19 layer VGG model \cite{simonyan2014very} in this work. We extract image features directly from the penultimate fully connected layer. The dimension of the image embedding is 2048 for ResNet152 and 4096 for VGG19. 

\vspace{0.1cm}
\textbf{Tag Representation:} We generate the feature representation of tags by summing over the Word2Vec \cite{mikolov2013efficient} embeddings of all tags associated with an image and then normalizing it by the number of tags. Averaged word vectors has been shown to be a strong feature for text in several tasks \cite{yu2014deep, kenter2015short, kenter2016siamese}. 
%especially when the order of tags is unknown.

\subsection{Train Joint Embedding with Ranking Loss} \label{vse}
We now describe the basic framework for learning joint image-sentence embedding based on bi-directional ranking loss. Many prior approaches have utilized pairwise ranking loss as the objective for learning joint embedding between visual input and textual input \cite{kiros2014unifying, zheng2017dual, wang2016learning, karpathy2014deep}. Specifically, these approaches minimize a hinge-based triplet ranking loss in order to maximize the similarity between an image embedding and corresponding text embedding and minimize similarity to all other non-matching ones.

Given an image feature representation $\overline{i}$~($\overline{i} \in \mathbb{R}^V$), the projection on the joint space can be derived as $i = W^{(i)}\overline{i}$ ($i \in \mathbb{R}^D$). Similarly, the projection of input text embedding $\overline{s}~ (\overline{s} \in \mathbb{R}^T)$ to joint space can be derived by $s = W^{(s)}\overline{s}~(s \in \mathbb{R}^D)$. Here, $W^{(i)} \in \mathbb{R}^{D \times V}$ is the transformation matrix that maps the visual content into the joint space and $D$ is the dimensionality of the space. In the same way, $W^{(s)}\in \mathbb{R}^{D \times T}$ maps input sentence embedding to the joint space. Given feature representation for words in a sentence, the sentence embedding $\overline{s}$ is found from the hidden state of the GRU. Here, given the feature representation of both images and corresponding text, the goal is to learn a joint embedding characterized by $\theta$ (i.e., $W^{(i)}$, $W^{(s)}$ and GRU weights) such that the image content and semantic content are projected into the joint space. Now, the image-sentence loss function $\mathcal{L}_{IS}$ can be written as, 
\begin{equation} \label{eq:equation1}
\begin{split}
\vspace{-0.2cm}
\mathcal{L}_{IS} = & \sum_{(i,s)}\Big\{\sum_{s^{-}}max \big[0, \Delta-f(i,s)+f(i,s^{-})\big] \\
 & \hspace{0.4cm} + \ \sum_{i^{-}} max \big[0, \Delta-f(s,i)+f(s,i^{-})\big]\Big\}
 \vspace{-0.1cm}
\end{split}
\end{equation}
where $s^{-}$ is a non-matching text embedding for image embedding $i$, and $s$ is the matching text embedding. This is similar for image embedding $i$ and non-matching image embedding $i^{-}$. $\Delta$ is the margin value for the ranking loss. The scoring function $f(i,s)$ measures the similarity between the images and text in the joint embedded space. In this work, we use cosine similarity in the representation space to calculate similarity, which is widely used in learning image-text embedding and shown to be very effective in many prior works \cite{zheng2017dual, kiros2014unifying, FaghriFKF17}. However, note that our approach does not depend on any particular choice of similarity function.

The first term in Eq.~(\ref{eq:equation1}) represents the sum over all non-matching text embedding $s^{-}$ which attempts to ensure that for each visual feature, corresponding/matching text features should be closer than non-matching ones in the joint space. Similarly, the second term attempts to ensure that text embedding that corresponds to the image embedding should be closer in the joint space to each other than non-matching image embeddings. 

Recently, focusing on hard-negatives has been shown to be effective in learning joint embeddings \cite{FaghriFKF17, zheng2017dual, schroff2015facenet, manmatha2017sampling}. Subsequently, the loss in Eq.~\ref{eq:equation1} is modified to focus on hard negatives (i.e., the negative closest to each positive $(i,s)$ pair) instead of sum over all negatives in the formulation. For a positive pair $(i,s)$, the hardest negative sample can be identified using $\hat{i} = \arg \max\limits_{i^{-}} f(s,i^{-})$ and $\hat{s} = \arg \max\limits_{s^{-}} f(i,s^{-})$. The loss function can be written as follows, 

\begin{equation} \label{eq:equation2}
\begin{split}
\vspace{-0.1cm}
\mathcal{L}_{IS} = & \sum_{(i,s)} \Big\{max \big[0, \ \Delta-f(i,s)+f(i,\hat{s})\big] \\
 & \hspace{0.4cm} + \  max \big[0, \ \Delta-f(s,i)+f(s,\hat{i})\big]\Big\}
\vspace{-0.1cm}
\end{split}
\end{equation}

We name Eq.~\ref{eq:equation1} as VSE loss and Eq.~\ref{eq:equation2} as VSEPP loss. We utilize both of these loss functions in evaluating our proposed approach.

\subsection{Training Joint Embedding with Web Data} \label{trimodal}

In this work, we try to utilize image-tag pairs from the web for improving joint embeddings trained using a clean dataset with images-sentence pairs. Our aim is to learn a good representation for image-text embedding that ideally ignores the data-dependent noise and generalizes well. Utilization of web data effectively increases the sample size used for training our model and can be considered as implicit data augmentation. However, it is not possible to directly update the embedding (Sec.~\ref{vse}) using image-tag pairs. GRU based approach is not suitable for representing tags since tags do not have any semantic context as in the sentences. 

Our task can also be considered from the perspective of learning with side or privileged information strategies \cite{vapnik2009new, sharmanska2013learning}, as in our case an additional tag modality is available at training time and we would like to utilize this extra information to train a stronger model. However, directly employing LUPI strategies are also not possible in our case as the training data do not provide information for all three modalities at the same time. The training datasets (e.g., MSCOCO, Flickr30K) provide only image-sentence pairs and do not provide tags. On the other hand, a web source usually provides images with tags, but no sentence descriptions. To bridge this gap, we propose a two-stage approach to train the joint image-text representation. In the first stage, we leverage the available clean image-text pairs from a dataset to learn an aligned representation that can be shared across three modalities (e.g., image, tag, text). In the second stage, we adapt the model trained in the first stage with web data.

\vspace{0.1cm}

\textbf{Stage I: Training Initial Embedding.} We leverage image-text pairs from an annotated dataset to learn a joint embedding for image, tags, and text. As tags are not available directly in the datasets, we consider nouns and verbs from the relevant sentence as dummy tags for an image (Fig.~\ref{fig:block_diagram}). For learning the shared representation, we combine the image-text ranking loss objective (Sec.~\ref{vse}), with image-tag ranking loss objective. We believe combining image-tag ranking loss objective provides a regularization effect in training that leads to more generalized image-text embedding.

Now the goal is to learn a joint embedding characterized by $\theta$ (i.e., $W^{(i)}$, $W^{(t)}$, $W^{(s)}$ and GRU weights) such that the image, sentence, and tags are projected into the joint space. Here, $W^{(t)}$ projects the representation of tags $\overline{t}$ on the joint space as, $t = W^{(t)}\overline{t}$. The resulting loss function can be written as follows,
\begin{equation} \label{eq:equation3}
\vspace{-0.0cm}
\mathcal{L} = \lambda_{1} \mathcal{L}_{IS} + \lambda_{2} \mathcal{L}_{IT}
 \vspace{-0.0cm}
\end{equation}
where, $\mathcal{L}_{IT}$ represent image-tag ranking loss objective, which is similar to image-sentence ranking loss objective $\mathcal{L}_{IS}$ in Sec.~\ref{vse}. Similar to VSEPP loss in Eq.~\ref{eq:equation2}, $\mathcal{L}_{IT}$ can be written as,

\begin{equation} \label{eq:equation4}
\begin{split}
\vspace{-0.1cm}
\mathcal{L}_{IT} = & \sum_{(i,t)} \Big\{max \big[0, \ \Delta-f(i,t)+f(i,\hat{t})\big] \\
 & \hspace{0.4cm} + \  max \big[0, \ \Delta-f(t,i)+f(t,\hat{i})\big]\Big\}
\vspace{-0.1cm}
\end{split}
\end{equation}
where for a positive image-tag pair $(i,t)$, the hardest negative sample tag representation can be identified as $\hat{t}$. Note that, all tags associated with an image are considered for generating tag representation in creating an image-tag pair rather than considering a single tag related to that image.
In Eq.~\ref{eq:equation3}, $\lambda_1$ and $\lambda_2$ are predefined weights for different losses. In the first training stage, both losses are used ($\lambda_1 = 1$ and $\lambda_2 = 1$) while in the second stage, the image-text loss is not used ($\lambda_1 = 0$ and $\lambda_2 = 1$).

\vspace{0.1cm}

\textbf{Stage II: Model Adaptation with Web Data.} After Stage I converges, we have a shared representation of image, sentence description and tags with a learned image-tag embedding model. In Stage II, we utilize weakly-annotated image-tags pairs from Flickr to update the previously learned embedding network using $\mathcal{L}_{IT}$ loss. This enables us to transfer knowledge from thousands of freely available weakly annotated images in learning the embedding. We utilize a smaller learning rate in Stage II, as the network achieves competitive performance after Stage I and tuning the embedding network with a high learning rate from weakly-annotated data may lead to catastrophic forgetting \cite{kemker2017measuring}. 

As web data is very prone to label noise, we found it is hard to learn good representation for our task in many cases. Hence, in Stage II, we adopt a curriculum learning-based strategy in training. Curriculum learning allows the model to learn from easier instances first so they can be used as building blocks to learn more complex ones, which leads to a better performance in the final task. It has been shown in many previous works that appropriate curriculum strategies guide the learner towards better local minima \cite{bengio2009curriculum}. Our idea is to gradually inject difficult information to the learner such that in the early stages of training, the network is presented with images related to frequently occurring concepts/keywords in the clean training set. Images related to rarely occurring concepts are presented at a later stage. Since the network trained in Stage I is more likely to have learned well about frequently occurring concepts, label noise is less likely to affect the network.

\section{Experiments} \label{experiments}
We perform experiments on two standard benchmark datasets with the main goal of analyzing the performance of different supervised methods by utilizing large scale web data using our curriculum guided webly supervised approach. Ideally, we would expect an improvement in performance irrespective of the loss function and features used to learn the embedding in Sec.~\ref{methods}. 

We first describe the details on the datasets and evaluation metric in Sec.~\ref{sec:data} and training details in Sec.~\ref{sec:training}. We report the results of different methods on MSCOCO dataset in Sec.~\ref{sec:coco} and results on Flickr30K dataset in Sec.~\ref{sec:flickr}. 

%Finally, we show some qualitative results in Sec.~\ref{sec:qual}.

\subsection{Datasets and Evaluation Metric}
\label{sec:data}

We present experiments on standard benchmark datasets for sentence-based image description: MSCOCO Dataset \cite{chen2015microsoft} and Flickr30K dataset \cite{plummer2015flickr30k} to evaluate the performance of our proposed framework.

\vspace{0.1cm}
\textbf{MSCOCO. } The MSCOCO is a large-scale sentence-based image description dataset. This is the largest image captioning dataset in terms of the number of sentences and the size of the vocabulary. This dataset contains around 123K images. Each image comes with 5 captions. Following \cite{karpathy2015deep}, we use the training, testing and validation split. In this split, the set contains 82,783 training images, 5000 validation images and 5000 test images. About 30K images were left out in this split. Some previous works utilize this images with for training to improve accuracy. We also report results using this images in training. In most of the previous works, the results are reported by averaging over 5 folds of 1K test images \cite{kiros2014unifying, wang2018learning, eisenschtat2017linking}.

\vspace{0.1cm}
\textbf{Flickr30K. } Flickr30K is another standard benchmark dataset for sentence-based image description. Flickr30K dataset has a standard 31,783 images and 158,915 English captions. Each image comes with 5 captions, annotated by AMT workers. In this work, we follow the dataset split provided in \cite{karpathy2015deep}. In this dataset split, the training set contains 29,000 images, validation set contains 1000  and test set contains 1000 images. 

\begin{table*}[t]
\vspace{0.1cm}
\centering
	\caption{Image-to-Text Retrieval Results on MSCOCO Dataset.}
	\vspace{-0.2cm}
	\includegraphics[width=0.80\textwidth]{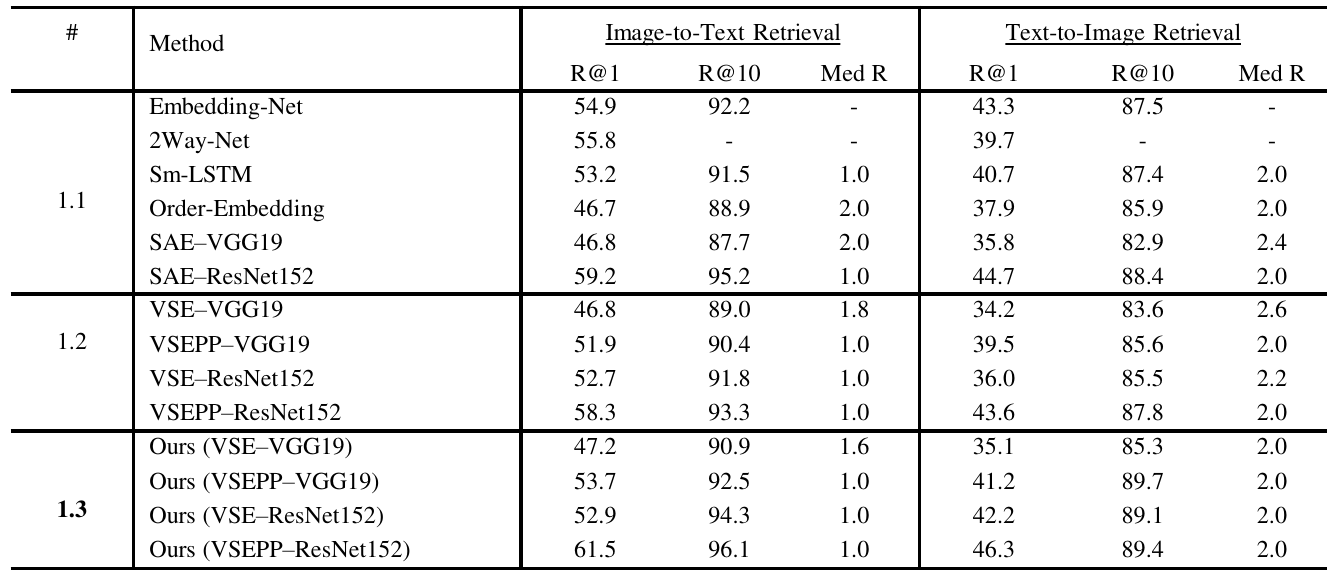}
	\label{tab:mscoco}
	\vspace{-0.3cm}
\end{table*}

\vspace{0.1cm}
\textbf{Web Image Collection. } We use photo-sharing website Flickr to retrieve web images with tags and use those images without any additional manual labeling. To collect images, we create a list of 1000 most occurring keywords in MSCOCO and Flickr30K dataset text descriptions and sort them in descending order based on frequency. We remove stop-words and group similar words together after performing lemmatization.  We then use this list of keywords to query Flickr and retrieve around 200 images per query, together with their tags. In this way, we collect about 210,000 images with tags. We only collect images having at least two English tags and we don't collect more than 5 images from a single owner. We also utilize first 5 tags to remove duplicate images. 

\vspace{0.1cm}
\textbf{Evaluation Metric. }
We use the standard evaluation criteria used in most prior work on image-text retrieval task~\cite{kiros2014unifying, FaghriFKF17, dong2016word2visualvec}. We measure rank-based performance by Recall at $K$ ($R@K$) and Median Rank($MedR$). $R@K$ calculates the percentage of test samples for which the correct result is ranked within the top-$K$ retrieved results to the query sample. We project sentences, tags, and images into the embedded space and calculate similarity scores. We report results of $R@1$ and $R@10$. Median Rank calculates the median of the ground-truth matches in the ranking results. 

\vspace{0.1cm}
\subsection{Training Details}
\label{sec:training}
We start training with a learning rate of 0.0002 and keep the learning rate fixed for 10 epochs. We then lower the learning rate by a factor of 10 every 10 epochs. We continue training Stage I for initial 20 epochs. Then we start updating the learned model in Stage I with web images in Stage II for another 20 epochs. The embedding networks are trained using ADAM optimizer \cite{kingma2014adam}. Gradients are clipped when the $L2$ norm of the gradients (for the entire layer) exceeds 2. We tried different values for margin $\Delta$ in training and empirically choose $\Delta$ as 0.2, which we found performed well consistently on the datasets. We evaluate the model on the validation set after every epoch. The best model is chosen based on the sum of recalls in the validation set to deal with the over-fitting issue. We use a batch-size of 128 in the experiment. We also tried with other mini-batch sizes of 32 and 64 but didn't notice significant impact on the performance. We used two Telsa K80 GPUs and implemented the network using PyTorch toolkit.

\subsection{Results on MSCOCO Dataset}
\label{sec:coco}

We report the result of testing on MSCOCO dataset \cite{lin2014microsoft} in Table~\ref{tab:mscoco}. To understand the effect of the proposed webly supervised approach, we divide the table in 3 rows (1.1-1.3). We compare our results with several representative image-text retrieval approaches, i.e., Embedding-Net \cite{wang2017learning}, 2Way-Net \cite{eisenschtat2017linking}, Sm-LSTM \cite{huang2017instance}, Order-Embedding \cite{vendrov2015order}, SAE \cite{gong2014improving}, VSE \cite{kiros2014unifying} and VSEPP \cite{FaghriFKF17}. For these approaches, we directly cite scores from respective papers when available and select the score of the best performing method if scores for multiple models are reported. 

In row-1.2, we report the results on applying two different variants of pair-wise ranking loss based baseline VSE and VSEPP with two different feature representation from \cite{FaghriFKF17}. VSE\cite{kiros2014unifying} is based on the basic triplet ranking loss similar to Eq.~\ref{eq:equation1} and VSEPP\cite{FaghriFKF17} is based on the loss function that emphasizes hard-negatives as shown in Eq.~\ref{eq:equation2}. We consider VSE and VSEPP loss based formulation as the main baseline for this work. Finally, in row-1.3, results using the proposed approach are reported. To enable a fair comparison, we apply our webly supervised method using the same VSE and VSEPP loss used by methods in row-1.2. 

\vspace{0.1cm}
\underline{\textit{Effect of Proposed Webly Supervised Training.}} For evaluating the impact of our approach, we compare results reported in row-1.2 and row-1.3. Our method utilizes the same loss functions and features used in row-1.2 for a fair comparison. 
From Table~\ref{tab:mscoco}, We observe that the proposed approach improves performance consistently in all the cases. For the retrieval task, the average performance increase in text-to-image retrieval is 7.5\% in R@1 and 3.2\% in R@10. 

We also compare the proposed approach with web supervised approach SAE\cite{gong2014improving} (reported in row-1.1). In this regard, we implement SAE based webly supervised approach following \cite{gong2014improving}. We use the same feature and VSEPP ranking loss for a fair comparison and follow the exact same settings for experiments. We observe that our approach consistently performs better.

\begin{table*}[t]
\vspace{0.1cm}
\centering
	\caption{Image-to-Text Retrieval Results on Flickr30K Dataset. }
	\vspace{-0.2cm}
	\includegraphics[width=0.81\textwidth]{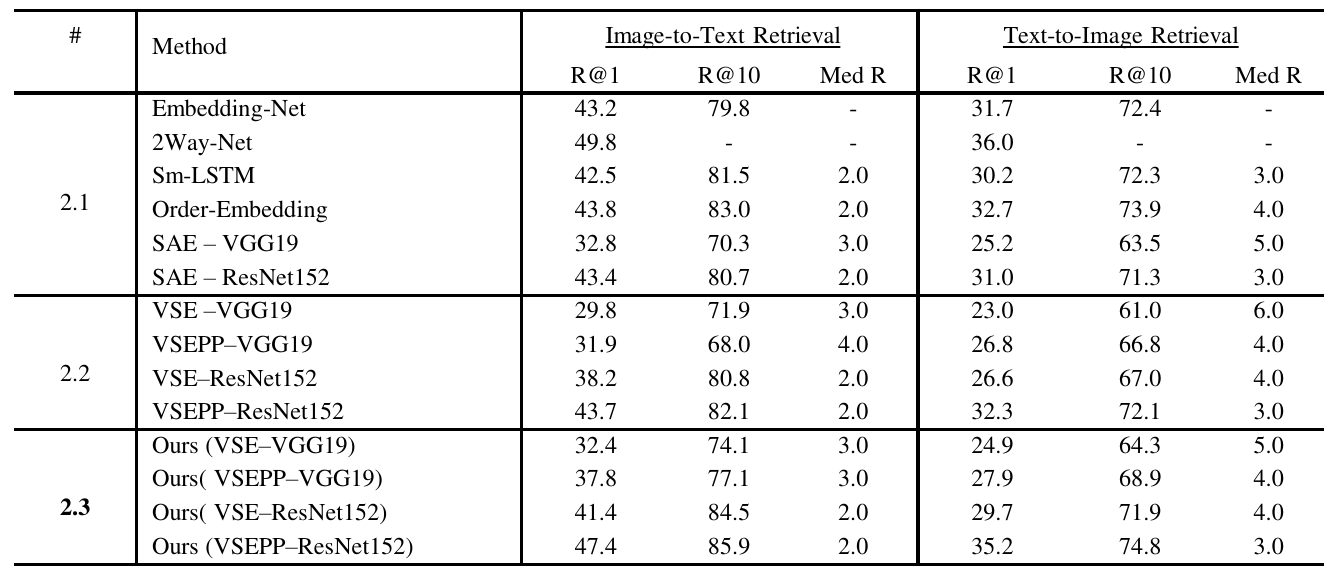}
	\vspace{0.1cm}
	\label{tab:f30k}
\end{table*}

%\vspace{4mm}

\begin{figure*}[t]
%\vspace{0.2cm}
\centering
	\includegraphics[width=0.90\textwidth]{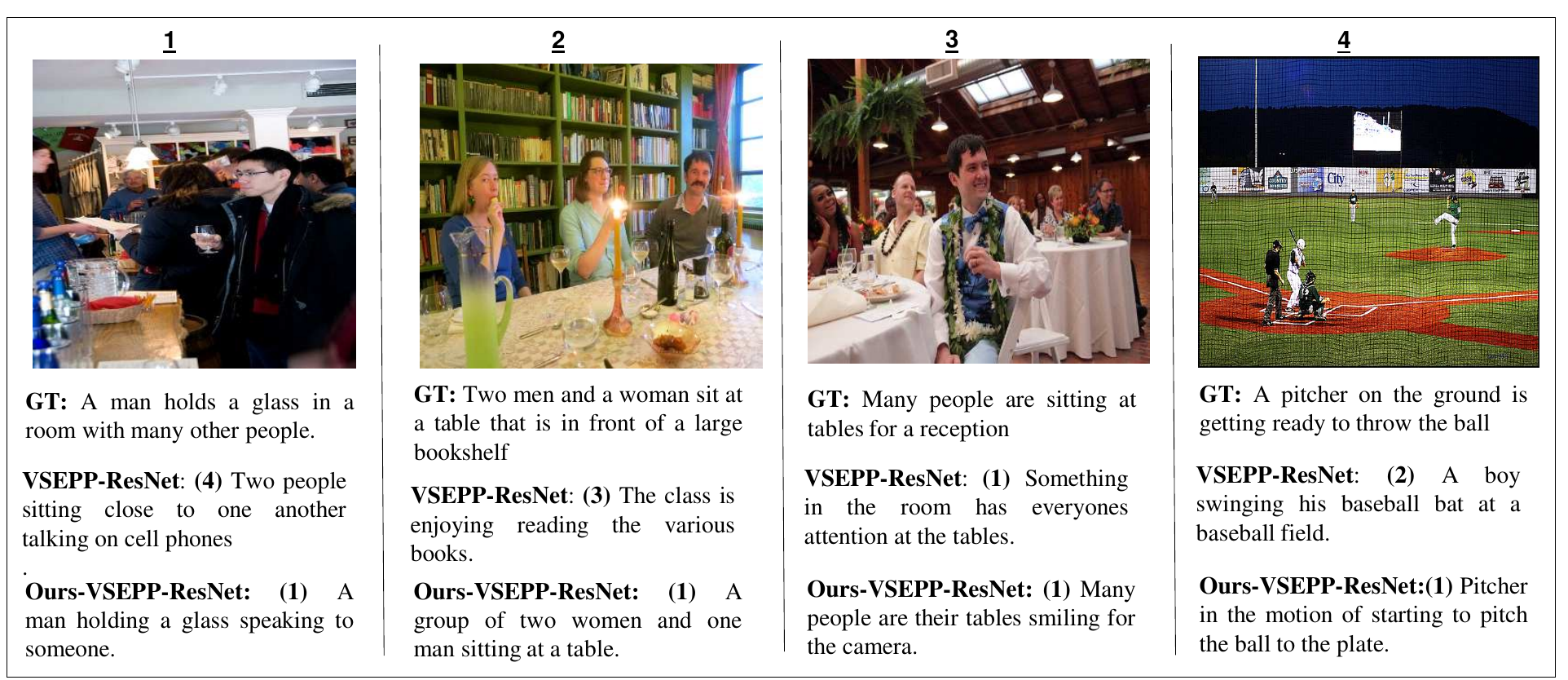}
	\vspace{-0.1cm}
	\caption{ Examples of 4 test images from Flickr30K dataset and the top 1 retrieved captions for our web supervised VSEPP-ResNet152 and standard VSEPP-ResNet as shown in Table.~\ref{tab:f30k}. The value in brackets is the rank of the highest ranked ground-truth caption in retrieval. Ground Truth (GT) is a sample from the ground-truth captions. Image 1,2 and 4 show a few examples where utilizing our approach helps to match the correct caption, compared to using the typical approach.
 }
	\label{fig:qualitative}
	\vspace{-0.2cm}
\end{figure*}

\vspace{0.1cm}
\underline{\textit{Effect of Loss Function.}} While evaluating the performance of different ranking loss, we observe that our webly supervised approach shows performance improvement for both VSE and VSEPP based formulation, and the performance improvement rate is similar for both VSE and VSEPP (See row-1.2 and row-1.3). Similar to the previous works \cite{FaghriFKF17, zheng2017dual}, we also find that methods using VSEPP loss perform better than VSE loss. We observe that in the image-to-text retrieval task, the performance improvement using VSEPP based formulation is higher and in the text-to-image retrieval task, the performance improvement for VSE based formulation is higher.

\vspace{0.1cm}
\underline{\textit{Effect of Feature.}} For evaluating the impact of different image feature in our web-supervised learning, we compare VGG19 feature based results with ResNet152 feature based results. We find consistent performance improvement using both VGG19 and ResNet152 feature. However, the performance improvement is slightly higher when ResNet152 feature is used. In image-to-text retrieval, the average performance improvement in R@1 using ResNet152 feature is 4\%, compared to 2.3\% using VGG19 feature. In the text-to-image retrieval task, the average performance improvement in R@1 using ResNet152 feature is 11.18\%, compared to 3.5\% using VGG19 feature.

Our webly supervised learning approach is agnostic to the choice loss function used for cross-modal feature fusion and we believe more sophisticated ones will only benefit our approach. We use two different variants of pairwise ranking loss (VSE and VSEPP) in the evaluation and observe that our approach improves the performance in both cases irrespective of the feature used to represent the images. 

\subsection{Results on Flickr30K Dataset}
\label{sec:flickr}
Table~\ref{tab:f30k} summarizes the results on Flickr30K dataset \cite{plummer2015flickr30k}. Similar to Table~\ref{tab:mscoco}, we divide the table into 3 rows (2.1-2.3) to understand the effect of the proposed approach compared to other approaches. From Table~\ref{tab:f30k}, we have the following key observations: (1) Similar to the results on MSCOCO dataset, our proposed approach consistently improves the performance of different supervised method (row-2.2 and row-2.3) in image-to-text retrieval by a margin of about 3\%-6\% in R@1 and 3\%-9\% in R@10. The maximum improvement of 6\%-9\% is observed in the VSEPP-VGG19 case while the least mean improvement of 4.8\% is observed in VSE-VGG19 case. (2) In text-to-image retrieval task, the average performance improvement using our webly-supervised approach are 2.25\% and 3.25\% in R@1 and R@10 respectively. These improvements once again show that learning by utilizing large scale web data covering a wide variety of concepts lead to a robust embedding for cross-modal retrieval tasks. In Fig.~\ref{fig:qualitative}, we show examples of few test images from Flickr30K dataset and the top 1 retrieved captions for the VSEPP-ResNet152 based formulations. 

%\subsection{Qualitative Results}
%\label{sec:qual}

\section{Conclusions}

In this work, we showed how to leverage web images with tags to assist training robust image-text embedding models for the target task of image-text retrieval that has limited labeled data. We attempt to address the challenge by proposing a two-stage approach that can augment a typical supervised pair-wise ranking loss based formulation with weakly-annotated web images to learn better image-text embedding. Our approach has benefits in both performance and scalability. Extensive experiments demonstrate that our approach significantly improves the performance in the image-text retrieval task in two benchmark datasets. Moving forward, we would like to improve our method by utilizing other types of metadata (e.g., social media groups, comments) while learning the multi-modal embedding. Furthermore, the objective of webly supervised learning may suffer when the amount of noisy tags associated with web images is unexpectedly high compared to clean relevant tags. In such cases, we plan to improve our method by designing loss functions or layers specific to noise reduction, providing a more principled way for learning the multi-modal embedding in presence of significant noise.

\vspace{0.2cm}
\hspace{-0.3cm}\textbf{Acknowledgement. } This work was partially supported by NSF grants IIS-1746031 and CNS-1544969. We thank Sujoy Paul for helpful suggestions and Victor Hill for setting up the computing infrastructure used in this work.
\balance
\bibliographystyle{ACM-Reference-Format}
\bibliography{bibliography}

\end{document}